\documentclass[prl,aps,twocolumn,floatfix,amsmath,amssymb,superscriptaddress,tightenlines]{revtex4}
\usepackage{graphicx}
\usepackage{epstopdf}
\usepackage{amsfonts}
\usepackage{bm}
\usepackage{color}
\usepackage{ulem}
\begin{document}

\newcommand{\be}{\begin{eqnarray}}
\newcommand{\ee}{\end{eqnarray}}

\date{\today}
 \title{Finite Size Scaling of Mutual Information: A Scalable Simulation}

\author{Roger G. Melko}
\affiliation{Department of Physics and Astronomy, University of Waterloo, Ontario, N2L 3G1, Canada} 

\author{Ann B. Kallin}
\affiliation{Department of Physics and Astronomy, University of Waterloo, Ontario, N2L 3G1, Canada} 

\author{Matthew B. Hastings}
\affiliation{Microsoft Research, Station Q, CNSI Building, University of California, Santa Barbara, CA, 93106}

\begin{abstract} 
We develop a quantum Monte Carlo procedure to compute the Renyi mutual information of an interacting quantum
many-body system at non-zero temperature.  Performing simulations on a spin-1/2 XXZ model, 
we observe 
that for a subregion of fixed size embedded in a system of size $L$, the mutual information converges at large $L$ to a limiting function which displays non-monotonic temperature behavior corresponding to the onset of correlations.
For a region of size $L/2$ embedded in a system of size $L$, the mutual information divided by $L$ converges to a limiting
function of temperature, with apparently nontrivial corrections near critical points.
\end{abstract}
\maketitle

Correlation functions have been an essential tool in determining properties of quantum systems.    However, in exotic phases, traditional correlation
functions may prove insufficient to capture hidden properties.
As a result, the Renyi entanglement entropies have attracted intense interest
recently for their abilities to deliver scaling terms with {\it universal} numbers, e.g.~at phase transitions \cite{Max} and in exotic
topologically ordered phases \cite{PI}, regardless of basis or choice of observable.
The generalized Renyi entanglement entropies, defined by
\begin{equation}
S_{n} (A) = \frac{1}{1-n} \ln \left[{ {\rm Tr}\big( \rho_A^{n} \big) }\right],
\label{Sn}
\end{equation}
(where $\rho_A$ is the reduced-density matrix of subregion $A$), have been successfully measured recently using $T=0$
projector QMC in the valence-bond basis, by employing the expectation value of a ``Swap'' operator \cite{QMC}.
However, to study more general physics using $S_n(A)$, one would like to develop a measure of entanglement entropy
applicable for models away from SU(2) symmetry and at $T>0$.

At non-zero temperature, the mutual information (MI) provides the appropriate analogue of the entanglement entropy, measuring information between one part of the system and another  \cite{wolf}.
In this paper, we develop a procedure for estimating the Renyi entropy in a finite-temperature QMC simulation, and
use it to calculate a Renyi MI, defined in analogy to the von Neumann MI: 
\be
I_n(A\colon \! \! B) = S_n(A) + S_n(B) - S_n(AB).
\ee
Here, $S_n(AB)$ is the $n$-th Renyi entropy for the whole system and $S_n(A),S_n(B)$ are the Renyi entropies for
subsystems $A$ and $B$ respectively.  At $T>0$, $I_n(A \colon \! \!  B)$ is expected to 
pick up both classical and quantum 
correlations, while at $T=0$, $I_n(A \colon \! \! B) = 2S_n(A) = 2S_n(B)$.

Stochastic Series Expansion (SSE) quantum Monte Carlo \cite{SSE1,SSE,DL} is a state-of-the-art finite-$T$ method, able to efficiently simulate systems of
particles with a large variety of U(1), SU(2) or SU(N) Hamiltonians.  In this paper, we use SSE to calculate the Renyi MI in the anisotropic XXZ model on a two-dimensional square lattice, with Hamiltonian,
\be
H = \sum_{\langle ij \rangle} \left({ \Delta S^z_i S^z_j + S^x_i S^x_j + S^y_i S^y_j }\right).
\label{Ham}
\ee
We verify our QMC procedure explicitly for small system sizes by showing that it reproduces $I_2(A \colon \! \! B)$ calculated with exact diagonalization (ED).  Then, by scaling the results
to large system sizes with QMC, we demonstrate that $I_2(A \colon \! \! B)$ effectively captures correlations at both $T>0$ and $T=0$.
We observe non-monotonic behavior in the MI, including a peak at high temperature, likely 
associated with a freezing out of classical correlations, followed by a rise at low temperature, due to the development of
long-range quantum entanglement.  Finally,  at the Ising transition for $\Delta > 1$, we observe novel scaling behavior of the MI, where a clear crossing appears at $T_c$ for different system sizes, possibly signaling a new type of universal scaling for this quantity.  

\begin{figure} {
\includegraphics[width=2.5 in]{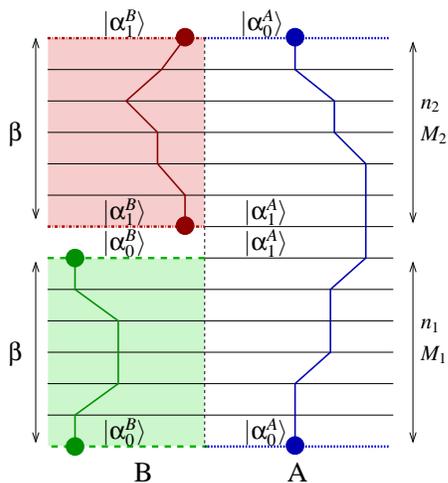} \caption{(color online) 
\label{Z2}
A schematic representation of the simulation cell for $Z[A,2,T]$ in finite-$T$ quantum Monte Carlo.  The one-dimensional basis is represented by the horizontal lines, and the vertical direction is the expansion, or imaginary time dimension.  World-lines (particle paths) are periodic in region $A$ over the time length of size $2 \beta$ (blue).  In the region $B$ (shaded), world lines are periodic in time $\beta$.}
} \end{figure}

{\it Finite-T QMC measurement of $S_2$.---}The {\it replica trick} \cite{Cardy} for calculating the generalized Renyi entropies has been used extensively 
in field-theory, both analytically \cite{Max} as well as in Monte Carlo simulations of lattice gauge theories \cite{BP,Naka}.  
The essential formulation is to consider the $d+1$ dimensional simulation cell with a 
modified topology -- that of an $n$-sheeted Riemann surface, with the region $A$ being periodic in $n \beta$ (in imaginary time), while the complement region $B$ is periodic in $\beta$.  Restricting to $n=2$
(see Fig.~\ref{Z2}), the partition function $Z[A,2,T]$ of the modified system is related to the trace of the 
reduced density matrix squared, 
\be
{\rm Tr} \rho^2_A = \frac{Z[A,2,T]}{Z(T)^2},
\ee
where $Z(T)$ is the partition function of the ``regular'' unmodified system at temperature $T$.

One immediately sees that the Renyi entropy will be accessible to any finite-T quantum Monte Carlo procedure that is based on a direct implementation of the modified partition function $Z[A,2,T]$.  At a given temperature, $S_2$ can be calculated for example through a thermodynamic integration from high temperatures, analogous to the calculation of free energy in classical systems.  In this case, integration of two
separate simulations are needed; one for $Z[A,2,T]$, and another for $Z(T)$:
\be
S_2(T) &=& -\ln Z[A,2,T] + 2 \ln Z(T) \nonumber \\ 
= &-& S_A(\beta=0) + \int_0^{\beta} \langle E \rangle_{A,\beta} d\beta \label{INT} \\
&+& 2S_0(\beta=0)  -  2 \int_0^{\beta} \langle E \rangle_{0,\beta} d\beta \nonumber.
\ee 
Here, the subscript $A,\beta$ refers to the energy estimator calculated at a given temperature $T=1/\beta$ in the $Z[A,2,T]$ simulation, and the subscript $0,\beta$ refers to $Z(T)$.  Note that the entropy $S_0(\beta=0) = N \ln(2)$ for an $N$-site spin 1/2 system.  For a system
with $N_A$ sites in subregion $A$, $S_A(\beta=0) = [N_A + 2(N-N_A)]\ln(2)$.  Simulation results used in this paper access $S_2(T)$ through Eq.~(\ref{INT}); other methods such as extended ensemble techniques can also be used.

{\it SSE Implementation.---}The above procedure can be implemented in any flavor of finite-T QMC based on the partition function, but we describe now a specific implementation in SSE QMC, referring the reader to the literature for the particular details and notation of the method \cite{DL}.  The SSE is based on a power-series expansion of the partition function.
The essential modification of the usual SSE formulation is simply the direct implementation of 
the modified boundary configuration for $Z[A,2,T]$, as described in the above illustration (Fig.~\ref{Z2}).
The $S^z$ basis state is decomposed into spins occurring in region $A$ and its complement $B$: $| \alpha \rangle = |\alpha^A \rangle | \alpha^B \rangle$. 
The SSE simulation cell is composed of this basis, and a list of operators that propagate $|\alpha \rangle$ through imaginary time.  To simulate $Z[A,2,T]$, two separate operator lists are employed, composed of $n_1$ and $n_2$ operators, which are each allowed to fluctuate independently.  Analogous to the typical SSE formulation, two separate cut-off variables are used, $M_1>n_1$ and $M_2>n_2$.  Three crucial changes are thus needed in the the SSE sampling of the operator list and basis, as described below. 

First, note that the probability to insert or remove a bond operator $H_b$ in the {\it diagonal update} is modified:
\be
P_{\rm add} &=& \frac{\beta N_b \langle \alpha_{\tau}| H_b | \alpha_{\tau+1} \rangle }{(M_\gamma - n_\gamma)}, \\
P_{\rm remove} &=& \frac{(M_\gamma - n_\gamma+1)}{\beta N_b \langle \alpha_{\tau}| H_b | \alpha_{\tau+1} \rangle },
\ee  
where $b$ is the lattice position of the bond operator (which can occur on $N_b$ nearest-neighbor bonds in 
Eq.~(\ref{Ham})), and $\tau$ is the ``time'' position in the expansion direction ($\tau \in [1,M_1+M_2]$.).  The index $\gamma $ has values 1 and 2 for $Z[A,2,T]$.  The two operators lists are sampled independently in the diagonal update \cite{DL}, thus their respective size ($n_1$ or $n_2$) is allowed to fluctuate independently.  
Since the inverse temperature $\beta$ occurs in each probability above, the total number of operators $n_1 + n_2$ is on average related to the total expansion direction $2 \beta$ (which also suggests the straightforward extension to the $n$-sheeted Riemann surface, where $\gamma = 1\cdots n$).

The second significant modification to the SSE sampling occurs in the loop update.  In fact, if one follows the standard two-step procedure in the loop update algorithm -- first constructing a linked list of connected ``vertices'' and second, performing loop updates in this linked list \cite{DL} -- then the modification reduces to a straight-forward reconnection of the topology of the linked list.  
Following intuition, the links between vertices are modified such that, if a spin in the vertex occurs in region $A$, connections are made as usual for the $Z(T)$ system with a periodic boundary at $2 \beta$ ($|\alpha^A_0 \rangle$ in Fig.~\ref{Z2}). 
For spins in region $B$, vertices containing operators in the list $n_1$ are connected only to themselves across the boundary $| \alpha^B_0 \rangle $ at $\beta$ (or, for vertices containing the operators in $n_2$, at  $| \alpha^B_1 \rangle$  -- Fig.~\ref{Z2}).  Remarkably, once this modified topology is implemented in the linked-list, the loop updates themselves are carried out as usual.  In particular, any vertex transition weights which are used in the usual simulation $Z(T)$ can be used without modification in the double-sheeted simulation $Z[A,2,T]$, including deterministic, heat-bath, or directed-loops \cite{DL}.

The final modification of note is the measurement of the expectation value of the energy in $Z[A,2,T]$, which must be re-derived along the lines of Ref.~\cite{SSE1}.  Then, one finds that the energy is related to the total number of operators in both lists,
\be
\langle E \rangle_{A,\beta} = \frac{\langle n_1+n_2 \rangle}{2 \beta}
\ee
where the additional factor of 2 in the denominator can be modified to $n$ for the generalized Renyi entropies.

\begin{figure} {
\includegraphics[width=3 in]{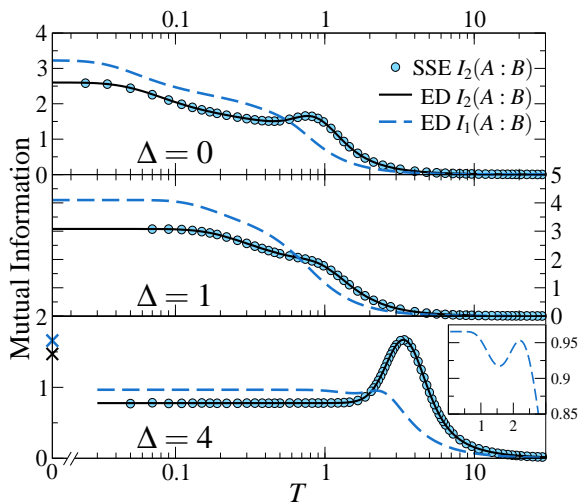} \caption{(color online) 
\label{ED}
Measurements of $I_1(A \colon \! \! B)$ (from ED) and $I_2(A \colon \! \! B)$ (from ED and SSE) for values of $\Delta =
0, 1, 4$ in Eq.~(\ref{Ham}).  In the lower plot ($\Delta = 4$) the exact $T=0$ values of $I_1(A \colon \! \! B)$ (higher) and $I_2(A \colon \! \! B)$ (lower) are shown.  The jump close to $T=0$ is caused by the near degeneracy of the lowest two energy levels of this system.  The inset shows the $I_1(A \colon \! \! B) $ curve on a linear scale, excluding the $T=0$ value, for $\Delta=4$.}
} \end{figure}

{\it Simulation results for MI.---}In Fig.~\ref{ED}, we show the MI for a system of size $4$-by-$4$ with Hamiltonian Eq.~(\ref{Ham}), where region $A$ is a square of size $3$-by-$3$.  The precise correspondence between SSE and ED data for $I_2$ clearly verifies the above method.  Note that with ED, we are also able
to compute $I_1$.  Two features are observed which deserve discussion below.

The first feature is the low temperature behavior of the MI for $\Delta>1$.  In the infinite system size limit, for any $\Delta > 1$ \cite{Doug}, 
the system spontaneously breaks a discrete Ising symmetry at low temperature.  As a result, for {\it finite} system sizes, the two lowest eigenvalues of the Hamiltonian, $E_0,E_1$ are almost degenerate ($E_1-E_0\approx 0.00009$ for
this system), with a gap $E_2-E_1$ to the rest of the spectrum.  For $E_1-E_0 \ll T \ll E_2-E_1$ the density matrix $\rho_{AB}$ for the whole system is close to a mixture of the two lowest states $\psi_0,\psi_1$:
\be
\rho_{AB}\approx (1/2)(|\psi_0\rangle\langle\psi_0| + |\psi_1\rangle\langle\psi_1|),
\ee
and so $S_1(AB)$ is very close to $\ln(2)$.  However, since $S_1(AB)=0$ at $T=0$,
the entropy $S_1(AB)$ jumps by $\approx \ln(2)$ over a very short interval of temperatures close to $T=0$.
In contrast, the entropies $S_1(A)$ and $S_1(B)$ are smooth at low temperatures.  Hence, for $\Delta>1$ the MI $S_1(A)+S_1(B)-S_1(AB)$ jumps by
approximately $\ln(2)$ at low temperatures due to breaking of the discrete symmetry -- as shown in Fig.~\ref{ED} for $\Delta=4$.
Thus, extracting the $T=0$ entanglement entropy from $T>0$ data is very difficult, requiring
the study of very low temperatures for $\Delta>1$.
For $\Delta\leq 1$, the splitting between low-lying eigenvalues is much larger, making it easier to extract the $T=0$ entanglement entropy, but the presence of low-lying
``tower of states" modes still suggests that a projector approach \cite{QMC} should be preferred to obtain
zero temperature data. 

A second interesting feature observed is a maximum in $I_2$ at finite-$T$ for $\Delta=0,4$.  For $\Delta=4$, this
is the global maximum, while for $\Delta=0$, the maximum is followed by a further increase at low temperature.  
Physically, we interpret these maxima as follows: at very high temperature, the system has no correlations and hence no MI.
As the temperature is initially lowered, the MI increases.  However, eventually the fluctuations in the system start to freeze out,
leading to a decrease in the MI.  Finally, at low temperature, long-range quantum entanglement begins to
appear, leading again to an increase in the MI.  
Note that one can also see a small maximum in the MI of $I_1$ for $\Delta=4$ in the inset at the bottom of Fig.~\ref{ED}

\begin{figure} {
\includegraphics[width=2.7 in]{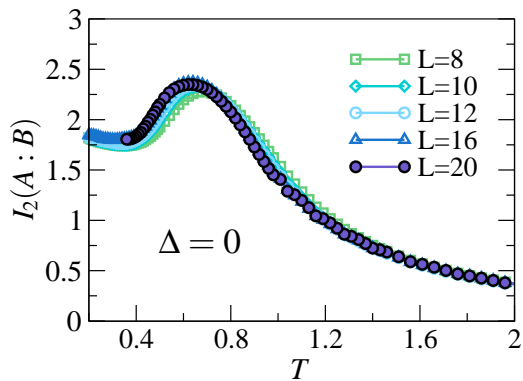} \caption{(color online) 
\label{conv}
Mutual information of the XY model for a square subregion $A$ of constant size $3 \times 3$.}
} \end{figure}

\begin{figure} {
\includegraphics[width=3.2 in]{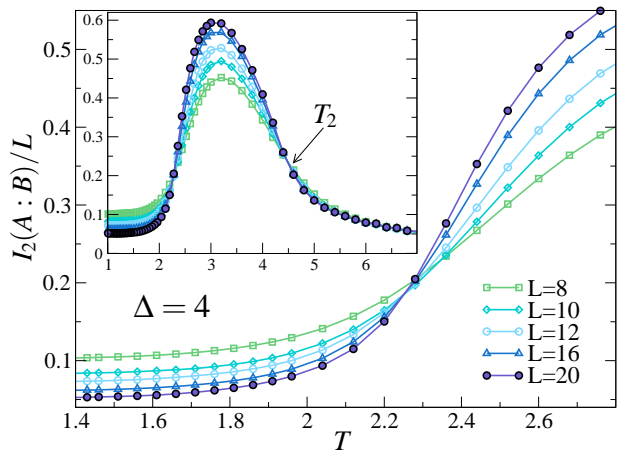} \caption{(color online) 
\label{Xing}
Mutual information scaled by the linear system size $L$, for the XXZ model with $\Delta = 4$ and a square subregion $A$ of size $L/2 \times L/2$.  The inset shows two crossing points for $I_2(A:B)/L$.  The main figure shows detail of the lower-$T$ crossing.}
} \end{figure}

{\it Scaling Behavior.---}In Fig.~\ref{conv} we plot the MI for a subregion of size $3$-by-$3$ embedded in a larger $L$-by-$L$ system for $\Delta=0$.  As $L$
increases, the curves approach a limiting function of temperature -- observed also for the other $\Delta$ values studied in this paper.
We also consider the case where region $A$ has size $L/2$-by-$L/2$ and the system has size $L$-by-$L$.  In Fig.~\ref{Xing}, we plot
$I_2(A \colon \! \! B)/L$ as a function of $L$ in the case of $\Delta=4$.  The inset shows $I_2(A \colon \! \! B)/L$ over a wide temperature range,
while the main figure is a close-up of data near $T=T_c \approx 2.25$ ($T_c$ was determined roughly by separate measurements of the
specific heat in $Z(T)$).  Dividing by $L$ shows the area law behavior, with the curves in the inset approaching
a limiting function of temperature.  This is an important feature of the MI: the difference
of entropies $S_2(A)+S_2(B)-S_2(AB)$ cancels terms proportional to volume, leaving a contribution proportional
to the surface area of $A$.

As one can see from the inset, the curves cross at two different temperatures.  The high temperature crossing, at $T_2\approx 4.45$,
can be understood simply on theoretical grounds.
For $T \gg T_c$, the correlations in the system are short-ranged, so we expect that
$I_2(A \colon \! \! B)=L f(T) + g(T) + {\cal O}(\exp(-L))$, where the term $g(T)$ is due to a correction to the
MI from the four corners of region $A$, while $Lf(T)$ represents a contribution from the edges of
region $A$ (indeed, we verified that this linear fit is very accurate for $T$ near $T_2$).
Then, the crossing corresponds to a change in sign of $g(T)$.

The lower temperature crossing is more remarkable, since it occurs at $T\approx T_c$.  Our explanation of the crossing
at high-$T$ relies on microscopic details (the particular sign of $g(T)$), so there does not at first seem to be
any reason for the low-$T$ crossing to occur near the critical point.  One might have dismissed this as a coincidence,
but we also observe for $\Delta=2$ the low-$T$ crossing again occurs at $T\approx T_c$.
To explain this behavior, we conjecture that the MI
as a function of temperature near $T_c$ has the form 
\be
I_2(A \colon \! \! B) = L f(T) + L^{\kappa} k((T-T_c)^{\nu} L)+g(T)+ ...
\ee
where $0<\kappa<1$ (possibly this term depends logarithmically on $L$ instead).  In this case, near the critical point the term $L^\kappa$
dominates the term $g(T)$ and the crossing is determined not by microscopic properties such as $g(T)$ but by universal properties.

Interestingly, the phenomenon of crossing points is also observed in the other models that we study.  For the Heisenberg case
($\Delta=1$), the upper crossing occurs at  $T_2 \approx 0.65$.  There exists crossings also below the peak in $I_2(A \colon \! \! B)/L$,
(around $T \approx 0.3$ for the $L$ studied), however the crossing point appears to decrease in temperature with increasing system size.  It would be interesting to determine if
$T_1\rightarrow 0$ as $\Delta\rightarrow 1$, following the trend in $T_c$. 
For the XY model, $T_2 \approx 0.85$, while the lower temperature crossing occurs around $T_1 \approx 0.41$ -- a slightly higher value than the Kosterlitz-Thouless temperature $T_{KT}=0.34303(8)$ \cite{TKT}.  A thorough study of the drift of this crossing point as a function of $L$, to see if it corresponds to $T_{KT}$ in the $L \rightarrow \infty$ limit would be interesting.

{\it Discussion.---}We have
presented a simulation method to measure mutual information at $T>0$, based on the Renyi entropy $S_n$, in interacting quantum many-body systems.  The procedure uses quantum Monte
Carlo based on a modified partition function, with efficient linear scaling in the spatial ($N$) and temporal ($n\beta$) simulation cell sizes.
Using SSE QMC, we examine the finite size scaling behavior of the $n=2$ mutual information 
in the spin 1/2 XXZ model with various anisotropies, and observe interesting non-monotonic behavior related to the onset of 
classical and quantum correlations.
The method has allowed us to extract novel behavior at critical points in the model, raising
interesting theoretical questions regarding the universal scaling properties of the mutual information, including the possibility of a new critical
exponent $\kappa$.  The generality of the idea for measuring mutual information, coupled with the versatility of finite-$T$ QMC methods
based on the partition function, should allow us to study a plethora of interesting questions in the immediate future, including scaling behavior in a quantum critical fan.

{\it Acknowledgments.---}The authors thank A. Sandvik, D.~Schwab, and A.~J.~Berlinsky for useful discussions, and the Boulder Summer School for Condensed Matter for hospitality.  This work was made possible by the
computing facilities of SHARCNET.  Support was provided by NSERC
of Canada (A.B.K. and R.G.M.).

\bibliography{Biblio}

\end{document}